# DICE AND PULSARS

*V.M. Kontorovich*

*Institute of Radio Astronomy NANU, Kharkov, Ukraine;*
vkont@ira.kharkov.ua

The pulse sequence is interpreted as a realization of a random electron discharge process in a vacuum gap over the polar cap (PC) —the open magnetic force line region on the neutron star surface. This point of view is illustrated by an example based on the dice. The generators of the random numbers are a cube and a coin. Throwing of dice and coin tossing determine the discharge places on the "light" and "dark" sides of PC, and correspondingly — the "shape" of the individual pulses and their statistical properties The physical mechanism giving such discharge scheme is shortly discussed. It may be a charge drained down from a sharp top of surface waves in a parallel electric field on the liquid PC surface of a neutron star.

## 1. INTRODUCTION

The known difficulties in interpretation of pulsar radio emission (see review and monograph [1], annotated bibliography [2], and monographs [3-5]) force us to recur once again to the physical processes that lead finally to the pulses registered by radio telescopes.

Although the "beacon" model and output of radiation from the region of open magnetic force lines over the polar cap of rotating neutron star are beyond any doubt [4,5], we have to accept that developed in the last quarter of the past century outstanding theories of complex processes in a magnetized plasma flowing down along the field lines (see references in [1-6]) do not give answer neither to a complex question about the bunching of electron-positron beams leading to coherent mechanism of radio emission with high radiation temperature, nor to a "simple" question — why any set of strongly distinguishing from each other pulses summarizes in an average profile reproduced with high accuracy and serving as "finger-prints" for each pulsar. We want to touch and partially discuss this issue not pretending to give complete answers to both questions. We start from the second question.

## 2. POWERFUL PULSES AND ABSENT PULSES

Curvature radiation mechanism proposes rather rigid link between the acceleration region of electron on the polar cap (PC) (and the force line of magnetic field) and region where the radiation emitted by the electron and its "descendants" is concentrated [4]. This is a result of relativistic aberration that associates each region on PC with the region on aperture — the transverse section of beacon beam — a sequence of "knots" with characteristic angular scale of order $1/\gamma$. (Lorentz factor of electrons accelerated in the vacuum gap over PC (Fig. 1) or created in the following $e \pm$ cascade is $\gamma \gg 1$). Thus, not interesting in a complex angular distribution of radiation we can "ascend" on the surface of PC and consider the located there acceleration regions which correspond to one or another observed pulse.

Let us make the simplest supposition that elementary "zones" — the regions on the PC surface from which the electrons tear themselves away (or "flow down" that more suits to the mechanism we will mention below) have ap-proximately the same surface for each pulse (more correctly, for the time interval corresponding to the acceleration and radiation processes forming the pulse).

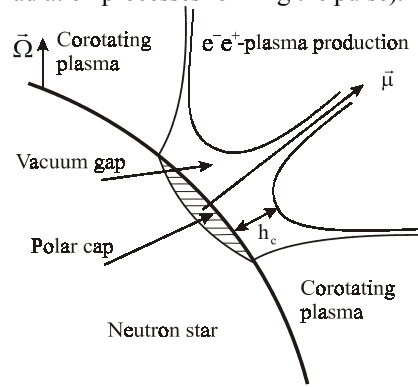

*Fig. 1.* *Polar cap and vacuum gap near the magnetic pole of a pulsar*

For typical pulsar both the time of plasma formation and flowing down along the field line to the light cylinder and the rotation period are of order of one second. Thus each following pulse is created by quite other particles accelerated in another accelerating process (in another discharge) and the other place of PC is linked with the other magnetosphere region. Here the places of acceleration are local regions on PC (the individual discharge places) from which the zones are constructed; we consider them having equal probability. In reality the regions of maximum electric field the probability of acceleration is bigger, and in the field line belt with greater curvature the radiation is more intensive. But at first step we will neglect these differences. So to each pulsar pulse a zone on the PC surface corresponds. (Moreover we do not suppose that the generation of the observed radiation arises just in these zones. It may arise in higher layers of the mentioned rigidly linked structure.) The stationary pulse is formed by summation of $N \gg 1$ pulses. That means that in the average pulse the radiation is already "gathered" (in the above mentioned sense) from the whole PC surface. The stationarity of PC properties determines the rigorous constancy of the average pulse form. At the same time the individuality of PC and magnetosphere (which among others play the role of transfer function to the observed pulse) in the region of open field lines for each pulsar procures the individuality of the average pulse for different pulsars. So, in each moment of the observed pulse creation not the whole aperture surface of beacon beam is illuminated but only $1/N$-th part of it. In general, we need to summarize about a thousand pulses to form the average one, because at the moment of pulse only the $1/N \sim 10^{-3}$ part of aperture will be illuminated. If each zone element enters some $(\beta)$ times during the average pulse formation, instantaneously shining zone will be not $1/N$ but $\beta/N$ part of the PC total surface. Here instantaneously radiating regions, generally speaking, have to represent an accidental picture of the spots depending on along what field line the primary electron beams are accelerating. In the earth telescope only part of aperture is seen which corresponds to "light" part of the PC surface. Thus the signal is received only from that part of zone which lies on the light part of the cap. If the zone totally lies on the "dark" part of the cap the signal will not be received by the telescope and the pulse is dropped out. Vice versa, if the active zone is fully placed on the light surface of PC, this case will correspond to the more powerful pulse. Alike with them in a few pulsars from time to time so cold "giant" pulses arise that they need special consideration.

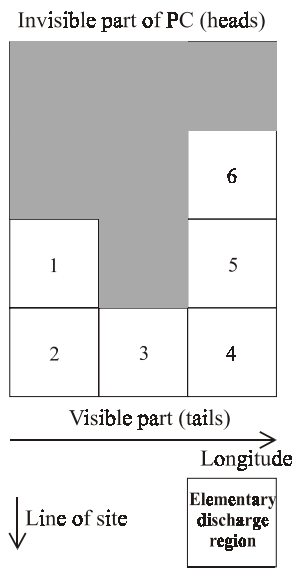

Invisible part of PC (heads)

Visible part (tails)

Longitude

Line of site

Elementary discharge region

**Fig. 2.** *Dark and light parts of Polar Cap for the 1st pulsar model. It is shown a symbolic longitude and direction of site line along which the signal is summarized forming an observed pulse. In real pulsars the connection between the place on PC and the pulse is very complicated due to the role of magnetosphere*

## 3. EXAMPLE BASED ON A DICE

We will illustrate the above using as random number generator a game cube with numbers from 1 to 6 and a coin, two sides of which correspond the "heads" (averse) and the "tail" (reverse). Let us assume that the "radiation" ("acceleration") zone in one pulse corresponds to two checks defined by two throws of the dice cube and the coin.

### 3.1. FIRST PULSAR MODEL

Here the tail corresponds to the light and the heads – to the dark part of the aperture that we will map in the form of the rectangle having 12 checks (Fig. 2) and stretched to 3 checks along the "longitude" and to 4 checks in the direction of the "line of sight". Let us agree to consider that the visible "signal" is summarized along the line of sight. Each signal will be summed from the two tossing of the cube and coin (Fig. 3), corresponding to the "pulse", stretched in the "pulse" along the longitude, the average form of which is represented as the light part of PC on the cube (Fig. 2). We have made not large number of the cube and coin tosses (16 tosses for the cube and 16 for the coin)

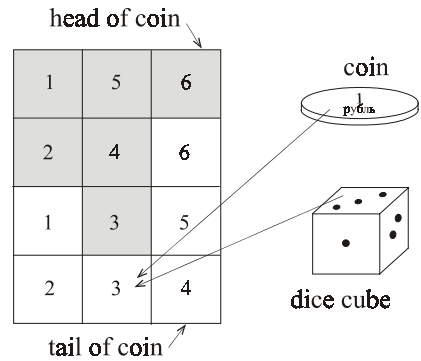

head of coin

| 1 | 5 | 6 |
| 2 | 4 | 6 |
| 1 | 3 | 5 |
| 2 | 3 | 4 |

tail of coin

coin

dice cube

**Fig. 3.** *The scheme of cube and coin tosses*

Of course this number of pulses is insufficient for the representative statistics but due to successful distribution in the test series is sufficient for an illustration. For each pulse the fallen numbers are denoted in Fig. 4 by the number of the dice cube and the side of the coin.

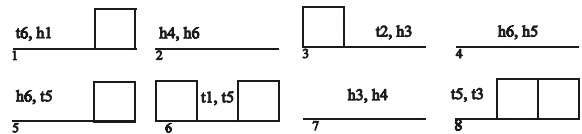

| t6, h1 | h4, h6 | t2, h3 | h6, h5 |
| 1 | 2 | 3 | 4 |
| h6, t5 | t1, t5 | t3, h4 | t5, t3 |
| 5 | 6 | 7 | 8 |

**Fig. 4.** *A series of 8 pulses for the first pulsar [7]. There are shown the results of cube throwing and coin tossing. The 6-th and 8-th pulses are powerful*

Pulses mapped in Fig. 4 correspond to the summing of signals from the light part of the PC along the line of sight. It is clear that 2nd, 4th and 7th pulses correspond to the absence of the pulse, 6th and 8th — to powerful pulses which in this model are only twice as "great" as the common ones. The number of absent pulses (=3) and powerful pulses (=2) with the adopted accuracy coincides. That is right because of equality of areas of dark and light parts of PC in this model, and the active zone has only two checks in it. It is remarkable that the result of 8 throws is the summarized pulse which practically coincides with the average one (the difference is only in one more check on the right side of the "cone"). But this is of course an accident (nice for us).

### 3.2. SECOND PULSAR MODEL

The formation of average pulse requires a significantly greater number of tosses. It can be easily verified with the change of the light and dark parts of the PC, considering "the 2nd pulsar" instead of the 1st one (Fig. 5). The corresponding pulses (from the former series of tests) are shown in Fig. 6. Here the 2nd, 4th and 7th pulses are powerful, 6th and 8th are the absent pulses.

Maybe the 7th pulse is a giant one. The summarized pulse in this version still differs essentially from the average one (after summarizing the 8 pulses).

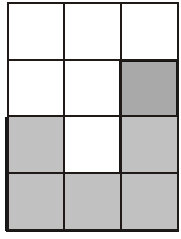

**Fig. 5.** *The second pulsar model. Rearranging the light and dark parts of PC we obtain the second pulsar model with small cone and big central part of the pulse*

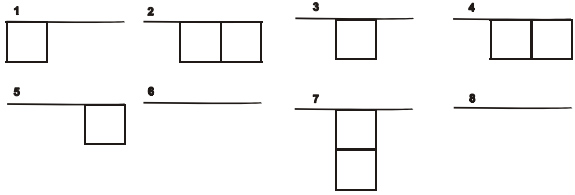

**Fig. 6.** *A series of the same 8 pulses for the second pulsar [7]. The 2-nd and 4-th pulses are powerful, the 7-th pulse may be a giant one*

The considered above case, in which the zone contains two summands ("discharges") and the light and dark parts of the PC equal each other, is not very representative. Playing with two dice cubes or even using the computer gives us greater opportunity. And we hope that the main idea is clear and return to our simplest game-model. In the pulse we may distinguish the contribution of "curvature cone" (the extreme columns) and "center". In the 1st pulsar (Fig. 2) the central radiation is weak and was observed only in 8th pulse. In the 2nd pulsar, vise versa, the cone is weaker and the center was observed in 3rd, 4th and 7th ("giant" pulse) pulses. Such differentiation of the pulse components is interesting for real pulsars also because of the fact that the radiation forming mechanism for the force line bundle near the magnetic pole, where the curvature radius is big and the radiation most likely caused by the inverse Compton effect or some beam mechanisms, differs in principle from the radiation mechanism on the "cone" where that is the curvature radiation which launches the following cascade.

## 4. QUANTITATIVE RELATIONS

Denote a number of powerful pulses during the time of formation of the pulse average profile through $N_P$ and a number of absent pulses for the same time through $N_A$. As far as each pulse is formed by $z$ independent discharges we have from the simple probabilistic reasons:

$$\frac{N_A}{N_P} = \left(\frac{D}{L}\right)^z, \quad \frac{N_P}{N} = \left(\frac{L}{L+D}\right)^z, \quad \frac{N_A}{N} = \left(\frac{D}{L+D}\right)^z. \quad (1)$$

Here $N$ is the number of pulses forming the average one, $D$ is the area of the dark part of PC, $L$ – the area of its light part. The quantities $N_A$, $N_P$, $N$ in the left-hand side of Eq. (1) are measurable. Expressing $D/L$ through the $N_A/N_P$ and substituting it in expression for $N_P/N$ we will receive the equation for parameter $z$, which, denoting $1/z = u$, is convenient to rewrite it in the form

$$1 + \left(\frac{N_A}{N_P}\right)^u = \left(\frac{N}{N_P}\right)^u. \quad (2)$$

Because $N > N_A + N_P$ the equation for $z$ always has a formal solution. For a numerical example see Fig. 7.

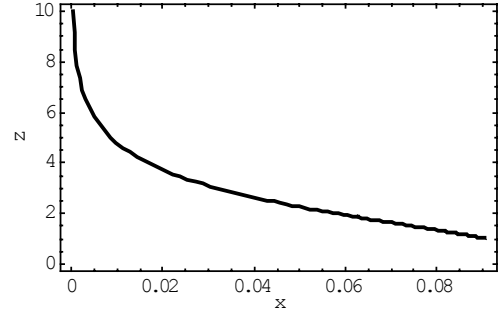

**Fig. 7.** *Numerical solution of equation (2) for* $N_A/N_P = 10$; $x \equiv N_P/N$

(Of course we have to take into account that $z$ is the integer number.) Note that in the particular case $D = L$ it has the explicit solution:

$$z = (\log N_P / N) / \log 2. \quad (3)$$

Knowing the whole surface of the PC $\Sigma$ we can find the surface $\sigma$ of elementary discharge:

$$\sigma = \Sigma / (z \cdot N / \beta). \quad (4)$$

Thus from the discussed scheme and with the knowledge of observed quantities $N$, $N_P$, $N_A$ and with the accurate enough estimation (from the geometrical considerations) of the PC surface $\Sigma$

$$\Sigma \approx \pi R_*^2 \cdot R_* / R_{LC}, \quad (5)$$

($R_*$ is the star radius and $R_{LC}$ is the light cylinder radius) we can find quantities $L$, $D$, $z$, and, if we know the parameter $\beta$, also the very important for physical theory surface $\sigma$. The quantity $\beta$ is order of unity and for enough rough estimation the exact value of it is not essential. (If may be found from the width of distribution for the average pulse).

If to consider in this scheme also the giant pulse [8-12] (Fig. 8) as a limit of localization of discharges on the PC, the corresponding probability of all $z$ discharges to hit into the area element $\sigma z$ on the light side of PC is

$$N_G / N = (\sigma z / \Sigma)^z. \quad (6)$$

The hitting of all $z$ discharges into the same area element $\sigma$ (if such process is possible) will give us

$$N_G / N = (\sigma / \Sigma)^z. \quad (7)$$

From other side we may try to estimate the quantity $z$ as the ratio of giant $A_G$ and common $A$ pulses amplitude as the generalization of case of 7-th pulse (Fig. 6) in our game:

$$z \approx A_G / A. \tag{8}$$

Taking into account also the possible nonlinear amplification [12] of the giant pulse, we have for the data from [10,11]:  $z \approx 10^2$,  $\sigma \approx 3 \cdot 10^2 \, \text{cm}^2$  (for  $(A_G / A)_{eff} \approx 10^2$,  $\Sigma \approx 10^8 \, \text{cm}^2$ ).

The useful for reconstruction of the PC parameters there could be also the pulse size distribution (may be non-Gaussian and having some features of Lévi statistics, see, for example, references in [13]).

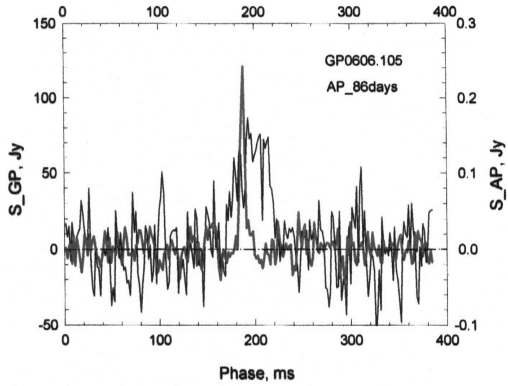

**Fig. 8.** *Giant pulse (left scale) in comparison with the common one (right scale). From [10, 11]*

## 5. PHYSICAL NATURE OF DISCHARGES ON THE SURFACE OF THE POLAR CAP

The fundamental is the presence of electric field, longitudinal to the magnetic one, which is discussed in a large number of works (see references in [1]). But a uniform ejection of charges from the surface would create a rather great charge density, screening of the field and disappearance of the vacuum gap, as it was mentioned in the literature [1]. Besides, further difficulties with bunching of radiating particles [1], that is necessary for explanation of the high brightness temperatures in the radio-frequency band make attractive the assumption that already from the edges of pulsar PC surface the electrons could flow down in the form of the filament currents of high density. These edges could be a part of neutron star crust but most likely they arise on the tops of gravitational or capillary surface waves in high magnetic [14] or/and electric field [15] in the case of melted liquid PC surface [1]. The arising currents are the kind of discharges resembling the lighting discharges.

The locality of these "thunderstorms" and quick flow down of plasma created by cascade beyond the limits of the light cylinder (the time of flow down is of order of 1 sec and is comparable to the period of pulsar rotation) have to lead to the tracery (openwork) plasma picture. In this picture the light clear spaces yawn permitting for both narrow directed needlelike radiation of relativistic particles and born by them the low frequency radiation to bypass the dense plasma regions and go out.

The "arch" (vault) itself of the vacuum gap is apparently not compact and continuous also, but represents changing from the pulse to pulse $e \pm$-plasma patterns located on some virtual surface. Such gap is partly similar to the perforated cylindrical waveguide (having the central cylindrical opening and a conical slot at the boundary of the region of closed magnetic field lines) and partly to the perforated resonator also. The "virtual" surfaces bounding them could be calculated in the frame of some models but in the process of radiation and frequency spectrum formation "instant" real arch surface plays its part, that make the problem very complicated. The resonator and waveguide are excited by discharges mentioned above and could be the sources of coherent radiation having high brightness temperature in the radio band. At the same time this low frequency radiation supplies quanta for inverse Compton scattering. This scattering could be an additional source of hard radiation with respect to curvature radiation. Perhaps it is the main mechanism of radiation in the region of small curvature of field lines near the pulsar magnetic axis and also in a narrow gap at the end of the region of closed field lines where the curvature radiation quanta do not penetrate. Besides, the Compton losses have to manifest themselves on the processes of energy accumulation by the primary electrons [16]. We hope to devote a separate paper to the discussion of some such possibilities.

## 6. CONCLUSION

The possibility to reduce the complex processes in the pulsar magnetosphere to the processes on the PC surface, we sure could be useful in the search of the clue to the intricate pulsar radiation. Of course, the simplest model presented above needs further improvement. Specifically the uniformity of PC surface does not take place. Rather it must resemble a target for shooting (in the first approximation with axial symmetry to the magnetic axis). The difference is that the circular zones around the center have the weight reflecting the angular distribution of the accelerating electric field (with the maximum in the center of PC) and the curvature of magnetic field lines (maximal at the edge of PC). Both these factors determine the intensity of radiation. From this point of view the powerful pulses could respond to the location of active zone in the belt corresponding to such maximal weight. The giant pulses may correspond to the localization of discharge in a small region of PC and the following switching on of nonlinear mechanisms in the magnetosphere. Then the simple relations discussed above will not take place. But the possibility arises to use more refined methods of geometrical probability theory. Note that nullings, or the frequent absences of pulses, from the presented here point of view may be the result of small probability for the active zone to hit in the light part of PC. This conforms to the small area of the PC light part and is determined by the geometry and orientation of PC. The alternative is the proximity to the "dead line", where the physical parameters of pulsar are such that the magnetosphere plasma cannot be generated and the radiation, as a result, cannot arise [17]. It is necessary to distinguish between these situations.

Some issues studied in this paper force us to return also to the problems discussed in the first reviews on the theme, such as [18,19], etc. Note that the proposed here model assumed from the beginning that the coordinates on the PC along the "longitude" and the "line of sight" could be introduced. Hence the PC could be mapped as it were in projection onto the coordinates of pulses. For real pulsar the linkage between physical coordinates on PC and pulse phase is determined by the transfer function of pulsar magnetosphere and can be found, in principle, in particular models of magnetosphere.

The author is grateful to those who participated in the discussion of this work at the seminar and Council of the Radio Astronomy Institute of NAS of Ukraine and also to O.V. Ulyanov for useful remarks.

## REFERENCES

1. V.S. Beskin. *Axially symmetric stationary flows in astrophysics.* M.: "Fizmatlit", 2006, 384 p.; Radio pulsars //*Usp.Fiz.Nauk.* 1999, v. 169, N11, p. 1169-1198 (in Russian).

2. S.B. Popov, M.E. Prohorov. *Astrophysics of single neutron stars: radio quiet neutron stars and magnitars.* M.: "GAISH MGU", 2003, 82 p. (in Russian).

3. I.F. Malov. *Radio Pulsars.* M.: "Nauka", 2004, 192 p. (in Russian).

4. F.G. Smith. *Pulsars.* M.: "Mir", 1979, 272 p. (in Russian).

5. R.N. Manchester and J.H. Taylor. *Pulsars.* M.: "Mir", 1980, 294 p. (in Russian).

6. V.M. Lipunov. *Astrophysics of neutron stars.* M.: "Nauka", 1987, 296 p. (in Russian).

7. V.M. Kontorovich. The Dice in connection with the Pulsars //*Radio Physics and Radio Astronomy.* 2006, v. 11, # 3, p. 308-314.

8. M.V. Popov, A.D. Kuzmin, O.M. Ulyanov, et al. Instantaneous spectrum of giant pulses of Crab pulsar from decimeter to decameter radio bands //*Astron. Zhurnal.* 2006, v. 83, # 7, p. 630-637.

9. O.M. Ulyanov, A.A. Zaharenko, A.A. Konovalenko, et al. Detection of individual pulses of pulsars in decametric band //*Radio Physics and Radio Astronomy.* 2006, v. 11, # 2, p. 113-133.

10. A.D. Kuzmin, A.A. Ershov. Detection of giant pulses of radio radiation of pulsar B0809+74 //*Astron.Lett.* 2006, v. 32, # 9, p. 650-654.

11. A.D. Kuzmin, Yu.P. Shitov. Pulsars – a new class of relativistic objects in the Universe //*Pushchino Radio Astronomy Observatory ASC Fian.* Puschino, 2006, p. 97-112.

12. S.A. Petrova. On the origin of giant pulses in radio pulsars //*Astron. Astrophys.* 2004, v. 424, p. 227-236.

13. A.V. Chechkin, V.Yu. Gonchar, J. Klafter, R. Metzler. Fundamentals of Lévy Flights //*Adv. Chem. Phys.* 2006, v. 133, Part B, p. 439-496.

14. M.I. Shliomis. Magnetic liquids //*Usp. Fiz. Nauk.* 1974, v. 112, N3, p.427-458 (in Russian).

15. N.M. Zubarev. Exact solutions of movement equations of liquid helium with free charged surface //*JETP.* 2002, v. 121, # 3, p. 624-636 (in Russian).

16. N.S. Kardashev, I.G. Mitrofanov, I.D. Novikov. Interaction of $e\pm$ with photons in magnetospheres of neutron stars //*JETP.* 1984, v. 61, # 6, p. 1113-1124 (in Russian).

17. E.M. Kantor and A.I. Tsygan. Switch out lines of pulsars in the case of dipole and asymmetric magnetic fields //*Astron. Zhurnal.* 2004, v. 81, # 6, p. 1130-1137.

18. V.L. Ginzburg. The Pulsars (Theoretical representations) //*Usp.Fiz.Nauk.* 1971, v. 103, N3, p. 393-429 (in Russian).

19. V.V. Usov. Pulsars //*Itogi nauki i tehniki, Space research.* VINITI, 1977, v. 9, p. 5-158 (in Russian).

## ИГРА В КОСТИ И ПУЛЬСАРЫ
### В.М. Конторович

Как следствие релятивистской аберрации и сверхсильного магнитного поля, имеется жесткая связь между областями ускорения (разряда) в магнитосфере пульсара и его наблюдаемым импульсом. Последовательность импульсов интерпретируется как реализация случайного процесса ускорения электронов в вакуумном зазоре над поверхностью полярной шапки (ПШ) – областью открытых силовых линий магнитного поля. Подход иллюстрируется примером, основанным на игре в кости, где генератором случайных чисел служат кубик и монетка, бросаниями которых определяются места разрядов на светлой и темной сторонах ПШ и соответственно индивидуальные импульсы. Обсуждается физический механизм, приводящий к подобной схеме ускорения. Им может быть стекание зарядов с заостренных вершин волн в параллельном электрическом поле на жидкой поверхности нейтронной звезды в области полярной шапки.

## ГРА В КОСТІ ТА ПУЛЬСАРИ
### В.М. Конторович

Як слідство релятивістської абераціі та надсильного магнітного поля, існує жорсткий зв'язок між місцями прискорення електронів у магнітосфері пульсара та імпульсом, що спостерігається. Послідовність імпульсів інтерпретується як реалізація випадкового процесу прискорення потоку електронів в вакуумному зазорі над поверхнею полярної шапки (ПШ) нейтронної зірки – областю відкритих силових ліній магнітного поля. Підхід ілюструється прикладом, що заснований на грі в кості, де генератором випадкових чисел є кубик та монетка. Їх киданням встановлюються місця розрядів на світлій або темній стороні полярної шапки і тим самим індивідуальні імпульси. Обговорюється фізичний механізм, що приводить до подібної схеми прискорення. Їм може бути стікання зарядів з загострених вершин хвиль в паралельнім магнітнім електричному полі на рідкій поверхні нейтронної зірки в області ПШ.